\documentclass[12pt]{article}
\usepackage{epsfig}
 
\renewcommand{\arraystretch}{1.3} 

\newcommand{\gtrsim}{\raisebox{-0.8mm}%
{\hspace{1mm}$\stackrel{>}{\sim}$\hspace{1mm}}}
\newcommand{\lessim}{\raisebox{-0.8mm}%
{\hspace{1mm}$\stackrel{<}{\sim}$\hspace{1mm}}}
\newcommand{\alphas}{\alpha_{\mrm{s}}}
\newcommand{\alphaem}{\alpha_{\mrm{em}}}
\newcommand{\ssintw}{\sin^2 \! \theta_W}
\newcommand{\scostw}{\cos^2 \! \theta_W}
\newcommand{\pT}{p_{\perp}}
\newcommand{\kT}{k_{\perp}}
\newcommand{\pTmin}{p_{\perp\mrm{min}}}
\newcommand{\br}[1]{\overline{#1}}
\newcommand{\pom}{\mrm{I}\!\mrm{P}}
\newcommand{\reg}{\mrm{I}\!\mrm{R}}
 
\newcommand{\mrm}[1]{\mathrm{#1}}
\renewcommand{\a}{{\mathrm a}}
\renewcommand{\b}{{\mathrm b}}
\renewcommand{\c}{{\mathrm c}}
\renewcommand{\d}{{\mathrm d}}
\newcommand{\e}{{\mathrm e}}
\newcommand{\f}{{\mathrm f}}
\newcommand{\g}{{\mathrm g}}
\newcommand{\hrm}{{\mathrm h}}
\newcommand{\lrm}{{\mathrm l}}
\newcommand{\n}{{\mathrm n}}
\newcommand{\p}{{\mathrm p}}
\newcommand{\q}{{\mathrm q}}
\newcommand{\s}{{\mathrm s}}
\renewcommand{\t}{{\mathrm t}}
\renewcommand{\u}{{\mathrm u}}
\newcommand{\A}{{\mathrm A}}
\newcommand{\B}{{\mathrm B}}
\newcommand{\D}{{\mathrm D}}
\newcommand{\F}{{\mathrm F}}
\renewcommand{\H}{{\mathrm H}}
\newcommand{\J}{{\mathrm J}}
\newcommand{\K}{{\mathrm K}}
\renewcommand{\L}{{\mathrm L}}
\newcommand{\Q}{{\mathrm Q}}
\newcommand{\R}{{\mathrm R}}
\newcommand{\W}{{\mathrm W}}
\newcommand{\Z}{{\mathrm Z}}
\newcommand{\bbar}{\overline{\mathrm b}}
\newcommand{\cbar}{\overline{\mathrm c}}
\newcommand{\dbar}{\overline{\mathrm d}}
\newcommand{\fbar}{\overline{\mathrm f}}
\newcommand{\pbar}{\overline{\mathrm p}}
\newcommand{\qbar}{\overline{\mathrm q}}
\newcommand{\sbar}{\overline{\mathrm s}}
\newcommand{\tbar}{\overline{\mathrm t}}
\newcommand{\ubar}{\overline{\mathrm u}}
\newcommand{\Fbar}{\overline{\mathrm F}}
\newcommand{\Qbar}{\overline{\mathrm Q}}
\newcommand{\Jpsi}{\mrm{J}/\psi}
\newcommand{\ee}{\e^+\e^-}
\newcommand{\ep}{\e\p}
\newcommand{\pp}{\p\p}
\newcommand{\pbarp}{\p\pbar}
\newcommand{\qqbar}{\q\qbar}
 
\newenvironment{Itemize}{\begin{list}{$\bullet$}%
{\setlength{\topsep}{0.2mm}\setlength{\partopsep}{0.2mm}%
\setlength{\itemsep}{0.2mm}\setlength{\parsep}{0.2mm}}}%
{\end{list}}
\newcounter{enumct}
\newenvironment{Enumerate}{\begin{list}{\arabic{enumct}.}%
{\usecounter{enumct}\setlength{\topsep}{0.2mm}%
\setlength{\partopsep}{0.2mm}\setlength{\itemsep}{0.2mm}%
\setlength{\parsep}{0.2mm}}}{\end{list}}
 
\renewcommand{\textfraction}{0.1}
\renewcommand{\topfraction}{0.9}
\renewcommand{\bottomfraction}{0.9}
\renewcommand{\floatpagefraction}{0.5}
 
\newlength{\captivewidth}
\newcommand{\captive}[1]{\rule{5mm}{0mm}%
\begin{minipage}{\captivewidth}%
\caption[small]{#1}\end{minipage}}
\newcommand{\onefigure}[2]{\begin{figure}[tb]\vspace{#1}
\captive{#2}\end{figure}}
 
\newlength{\abstwidth}
\begin{document}
 
\sloppy
 
\pagestyle{empty}
 
\begin{flushright}
CERN--TH/96--119  \\
LU TP 96--13 \\
May 1996
\end{flushright}
 
\vspace{\fill}
 
\begin{center}
{\LARGE\bf A scenario for high-energy \boldmath $\gamma\gamma$ %
interactions}\\[10mm]
{\Large Gerhard A. Schuler$^a$} \\[3mm]
{\it Theory Division, CERN,} \\[1mm]
{\it CH-1211 Geneva 23, Switzerland}\\[1mm]
{ E-mail: schulerg@cernvm.cern.ch}\\[2ex]
{\large and} \\[2ex]
{\Large Torbj\"orn Sj\"ostrand} \\[3mm]
{\it Department of Theoretical Physics,}\\[1mm]
{\it University of Lund, Lund, Sweden}\\[1mm]
{ E-mail: torbjorn@thep.lu.se}
\end{center}
 
\vspace{\fill}
 
\begin{center}
{\bf Abstract}\\[2ex]
\begin{minipage}{\abstwidth}
A real photon has a complicated nature, whereby it may remain 
unresolved or fluctuate into a vector meson or a perturbative 
$\q\qbar$ pair. In $\gamma\gamma$ events, this gives 
three by three combinations of the nature of the two incoming 
photons, and thus six distinct event classes. The properties
of these classes are partly constrained by the choices already 
made in our related $\gamma\p$ model. It is therefore possible 
to predict the energy-dependence of the cross section for each 
of the six components separately. The total cross section  gives 
support to the idea that a simple factorized ansatz with a pomeron 
and a reggeon term can be a good approximation. Event properties 
undergo a stepwise evolution from $\p\p$ to $\gamma\p$ to 
$\gamma\gamma$ events, with larger charged multiplicity, more 
transverse energy flow and a higher jet rate in the latter process.
\end{minipage}
\end{center}
 
\vspace{\fill}
\noindent
\rule{60mm}{0.4mm}
 
\vspace{1mm} \noindent
${}^a$ Heisenberg Fellow.
 
\vspace{10mm}\noindent
CERN--TH/96--119
 
\clearpage
\pagestyle{plain}
\setcounter{page}{1}
 
\section{Introduction}

There are many reasons for being interested in $\gamma\gamma$ physics.
The collision between two photons provides the richest
spectrum of (leading-order) processes that is available for any choice 
of two incoming elementary particles. For instance, since the photon 
has a hadronic component, all of hadronic physics is contained as a 
subset of the possibilities. Additionally, the photon can appear
as an unresolved particle or as a perturbative $\q\qbar$ fluctuation,
giving a host of possible further interaction processes. The relative
amount of these components and their respective properties are not
fully understood today. A correct description of the components of 
the total $\gamma\gamma$ cross section and the related event shapes
therefore is the ultimate challenge of `minimum-bias' physics.
Specific issues include the description of the photon wave function,
duality between perturbative and nonperturbative descriptions of the
resolved photon, the r\^ole of multiple parton--parton interactions
and the related minijet phenomenology and eikonalization of the total 
cross section, the transition between soft and hard physics, the 
transition between the real photon and the virtual one, and so on.

In addition to the direct reasons, there are also indirect ones.
The process $\ee \to \ee\gamma\gamma \to \ee X$ will be a main one	
at LEP 2 and future linear $\ee$ colliders. Therefore, 
$\gamma\gamma$ events are always going to give a non-negligible 
background to whatever other physics one is interested in.
The devising of efficient analysis strategies must be based on
a good understanding of $\gamma\gamma$ physics.

The study of $\gamma\gamma$ physics has a long history, and it is 
not our intention here to give a complete list of references. 
Many topics have been covered by contributions to past workshops 
\cite{workshops}. In recent years HERA has been providing rich 
information on the related $\gamma\p$ processes, and thereby 
stimulating the whole field \cite{HERAwork}. Further developments 
can be expected here. Currently $\gamma\gamma$ interest is focussed 
on LEP~2 \cite{LEP2workshop}. Already LEP~1.5 has provided ample 
reminder of the important r\^ole of $\gamma\gamma$ processes when 
away from the $\Z^0$ pole. LEP~2 brings the promise of a large event
rate at reasonably large $\gamma\gamma$ energies. In the future, 
the laser backscattering option of linear $\ee$ colliders offers the 
promise of obtaining a large rate of very high-energy $\gamma\gamma$ 
interactions, typically at up to 70\% of the energy of the 
corresponding $\ee$ collisions. Then the two aspects above come 
together in force, both with new chances to understand photon
interactions and new challenges to eliminate the $\gamma\gamma$
background to other processes.  

The starting point for the current paper is our model for $\gamma\p$ 
physics \cite{gammap}. Many of the basic assumptions can be taken
over in an (almost) minimal fashion, while further new ones appear. 
Based on the experience from HERA, some old assumptions can be sharpened 
and further developed. LEP~2 and future linear colliders will allow
new tests to be carried out. Our recent studies on the parton 
distributions of the photon \cite{ourpdf} are parts of the same 
physics program, and provide further important building blocks for 
the current study. In this paper we also emphasize the 
gradual evolution from $\p\p$ to $\gamma\p$ to $\gamma\gamma$ events,
that allows some cross-checks to be carried out systematically.
Parts of this work has already been presented in a preliminary
form at workshops \cite{previous,previoustwo}. 

No model exists in a vacuum. For the approach we are going to take, 
one important line of work is the subdivision of photon interactions 
by the nature of the photon \cite{PWZ}. Minijet phenomenology has 
attracted much attention in recent years \cite{minijet}. 
Other related works will appear as we go along.  
However, none of these approaches attempts to give a complete 
description of $\gamma\gamma$ cross sections and event properties,
but only concentrate on specific topics. Here we will try to be 
more ambitious, and really provide all the necessary aspects in one
single framework. The only other work with a somewhat similar
global scope is the recent studies within the context of 
dual topological unitarization \cite{DTU}.

Some main areas are still left out of our description. In all that
follows, both incoming photons are assumed to be on the mass shell;
further issues need to be addressed when either photon or both of them
are virtual. The issue of the eikonalization of
the anomalous and direct components of the photon wave function will 
be partly deferred to future studies; the evidence for some form of 
eikonalization will become apparent as we go along.
Finally, for reasons of clarity, we 
restrict ourselves to discussing what happens in the collision 
between two photons of given momenta. The addition of photon flux 
factors \cite{LEP2workshop} complicates the picture, but does not 
add anything fundamentally new.

Section 2 contains a description of the photon wave function and 
$\gamma\gamma$ event classes, section 3 of total and partial
cross sections, section 4 of event properties and section 5 gives
a summary and outlook.

\section{The Photon Wave Function and Event Classes}

To first approximation, the photon is a point-like particle. However,
quantum mechanically, it may fluctuate into a (charged) 
fermion--antifermion pair. The fluctuations 
$\gamma \leftrightarrow \q\qbar$ are of special interest to us,
since such fluctuations can interact strongly and therefore turn 
out to be responsible for the major part of the $\gamma\p$ and 
$\gamma\gamma$ total cross sections, as we shall see. On the other 
hand, the fluctuations into a lepton pair are uninteresting, since 
such states do not undergo strong interactions to leading order, and
therefore contribute negligibly to total hadronic cross sections. The
leptonic fluctuations are perturbatively calculable, with an
infrared cut-off provided by the lepton mass itself. Not so for
quark pairs, where low-virtuality fluctuations enter a domain of 
non-perturbative QCD physics. It is therefore customary to split
the spectrum of fluctuations into a low-virtuality and a high-virtuality
part. The former part can be approximated by a sum over low-mass 
vector-meson states, customarily (but not necessarily) restricted 
to the lowest-lying vector multiplet. Phenomenologically, this 
Vector Meson Dominance (VMD) ansatz turns out to be very successful in
describing a host of data. The high-virtuality part, on the other hand, 
should be in a perturbatively calculable domain. 

In total, the photon wave function can then be written as \cite{gammap}
\begin{equation}
|\gamma\rangle = c_{\mrm{bare}} |\gamma_{\mrm{bare}}\rangle +
\sum_{V = \rho^0, \omega, \phi, \Jpsi} c_V |V\rangle +
\sum_{\q = \u, \d, \s, \c, \b} c_{\q} |\q\qbar\rangle +
\sum_{\ell = \e, \mu, \tau} c_{\ell} |\ell^+\ell^-\rangle 
\label{gammawavefunction}
\end{equation} 
(neglecting the small contribution from $\Upsilon$). In general, the 
coefficients $c_i$ depend on the scale $\mu$ used to probe the photon.
Thus $c_{\ell}^2 \approx (\alphaem/2\pi)(2/3) \ln(\mu^2/m_{\ell}^2)$. 
Introducing a cut-off parameter $k_0$ to separate the low- and 
high-virtuality parts of the $\q\qbar$ fluctuations, one similarly 
obtains $c_{\q}^2 \approx (\alphaem/2\pi) 2e_{\q}^2 \ln(\mu^2/k_0^2)$.
Since each $\q\qbar$ fluctuation is characterized by some virtuality 
or transverse momentum scale $k$, the notation in 
eq.~(\ref{gammawavefunction}) should really be viewed as shorthand,
where the full expression is obtained by
\begin{equation}
c_{\q} |\q\qbar\rangle \longmapsto \frac{\alphaem}{2\pi} \, 
2e_{\q}^2 \, \int_{k_0^2}^{\mu^2} \frac{\d k^2}{k^2} \,
|\q\qbar;k^2\rangle ~,  
\end{equation}
and correspondingly for the lepton component. This is the form 
assumed in the following.
The VMD part corresponds to the range of $\q\qbar$ fluctuations below
$k_0$ and is thus $\mu$-independent (assuming $\mu > k_0$). 
In conventional notation
$c_V^2 = 4\pi\alphaem/f_V^2$, with $f_V^2/4\pi$ determined from data
to be 2.20 for $\rho^0$, 23.6 for $\omega$, 18.4 for $\phi$ and 
11.5 for $\Jpsi$ \cite{Baur}. Finally, $c_{\mrm{bare}}$ is given by
unitarity: $c_{\mrm{bare}}^2 \equiv Z_3 = 1 - \sum c_V^2 -
\sum c_{\q}^2 - \sum c_{\ell}^2$. In practice, $c_{\mrm{bare}}$ is
always close to unity. Usually the probing scale $\mu$ is taken to be 
the transverse momentum of a $2 \to 2$ parton-level process. Our fitted
value $k_0 \approx 0.5$ GeV \cite{gammap} then sets the minimum 
transverse momentum of a perturbative branching $\gamma \to \q\qbar$.

In part, $k_0$ is an unphysical parameter, and one would expect a
continuity under reasonably variations of it. That is, if the VMD sum 
were to be extended beyond the lowest-lying vector mesons to also
include higher resonances, it should be possible to compensate this
by using a correspondingly higher $k_0$ cut-off for the continuous
perturbative spectrum. This may provide some guidelines when exploring 
the physics implications of the ansatz above. The VMD--perturbative
state duality has its limits, however. Higher excitations of vector mesons
have larger wave-function radii than the lowest-lying states when each
is produced `on shell' in the time-like region (roughly $r \propto m$), 
while the uncertainty relation gives smaller radii for the 
higher-virtuality components of the real photon wave function
($r \propto 1/m$). Correspondingly, the contributions of $\rho^0$,
$\omega$ and $\phi$ could not well be described by perturbation
theory alone: the $c_V$ parameters are related to the absolute rates of 
the elastic processes $\gamma \p \to V \p$, and here the observed 
relation between $\rho^0$ and $\omega$ production is in
agreement with the $9:1$ VMD expectations of coherent vector meson 
wave-functions, while incoherent interactions of perturbative
components $|\u\ubar\rangle$ and $|\d\dbar\rangle$ would have lead to 
equal production of $\rho^0$ and $\omega$.

The subdivision of the above photon wave function corresponds to the 
existence of three main event classes in $\gamma\p$ events, 
cf. Fig.~\ref{fig1}:
\begin{Enumerate}
\item The VMD processes, where the photon turns into a vector meson
before the interaction, and therefore all processes
allowed in hadronic physics may occur. This includes elastic and 
diffractive scattering as well as low-$\pT$ and high-$\pT$ 
non-diffractive events.
\item The direct processes, where a bare photon interacts with a 
parton from the proton.
\item The anomalous processes, where the photon perturbatively branches
into a $\q\qbar$ pair, and one of these (or a daughter parton thereof)
interacts with a parton from the proton. 
\end{Enumerate}
All three processes are of $O(\alphaem)$. However, in the direct 
contribution the `parton' distribution of the photon is of $O(1)$ and 
the hard scattering matrix elements of $O(\alphaem)$, while the 
opposite holds for the VMD and the anomalous processes. 
As we already noted, the $\ell^+\ell^-$ fluctuations are not
interesting, and there is thus no class associated with them.

The above subdivision is not unique, or even the conventional one. More
common is to lump the jet production processes of VMD and anomalous
into a class called resolved photons. The remaining `soft-VMD' class 
is then 
defined as not having any jet production at all, but only consisting
of low-$\pT$ events. We find such a subdivision 
counterproductive, since it is then not possible to think of the 
VMD class as being a scaled-down version (by a factor $c_V^2$) of
ordinary hadronic processes --- remember that normal hadronic
collisions {\em do} contain jets part of the time. 

In a complete framework, there would be no sharp borders between the
three above classes, but rather fairly smooth transition regions that 
interpolate between the extreme behaviours. However, at our current
level of understanding, we do not know how to do this, and therefore 
push our ignorance into parameters such as the $k_0$ scale and the 
$f_V^2/4\pi$ couplings. From a practical point of view, the sharp 
borders on the parton level are smeared out by parton showers and
hadronization. Any Monte Carlo event sample intended to catch a border 
region would actually consist of a mixture of the three 
extreme scenarios, and therefore indeed be intermediate. 
This issue is discussed further in section~\ref{subsectdiranom}.

The difference between the three classes is easily seen in terms 
of the beam jet structure. The incoming proton always gives a beam jet
containing the partons of the proton that did not interact. On the
photon side, the direct processes do not give a beam jet at all, since
all the energy of the photon is involved in the hard interaction. The
VMD ones (leaving aside the elastic and diffractive subprocesses for the 
moment) give a beam remnant just like the proton, with a `primordial
$k_{\perp}$' smearing of typically up to half a GeV. The anomalous 
processes give a beam remnant produced by the $\gamma \to \q\qbar$
branching, with a transverse momentum going from $k_0$ upwards. 
Thus the transition from VMD to anomalous should be rather smooth.

A generalization of the above picture to $\gamma\gamma$ events is 
obtained by noting that each of the two incoming photons is
described by a wave function of the type given in 
eq.~(\ref{gammawavefunction}). In total, there are therefore 
three times three
event classes. By symmetry, the `off-diagonal' combinations appear
pairwise, so the number of distinct classes is only six.
These are, cf. Fig.~\ref{fig2}:
\begin{Enumerate}
\item VMD$\times$VMD: both photons turn into hadrons, and the processes
are therefore the same as allowed in hadron--hadron collisions.
\item VMD$\times$direct: a bare photon interacts with the partons of the
VMD photon.
\item VMD$\times$anomalous: the anomalous photon perturbatively 
branches into a $\q\qbar$ pair, and one of these (or a daughter parton 
thereof) interacts with a parton from the VMD photon.
\item Direct$\times$direct: the two photons directly give a quark pair,
$\gamma\gamma \to \q\qbar$. Also lepton pair production is allowed,
$\gamma\gamma \to \ell^+\ell^-$, but will not be considered by us.
\item Direct$\times$anomalous: the anomalous photon perturbatively 
branches into a $\q\qbar$ pair, and one of these (or a daughter parton 
thereof) directly interacts with the other photon. 
\item Anomalous$\times$anomalous: both photons perturbatively branch 
into $\q\qbar$ pairs, and subsequently one parton from each photon 
undergoes a hard interaction.
\end{Enumerate}
The first three classes above are pretty much the same as the three 
classes allowed in $\gamma\p$ events, since the interactions of a VMD
photon and those of a proton are about the same.

The main parton-level processes that occur in the six classes are:
\begin{Itemize}
\item The `direct' processes $\gamma\gamma \to \q\qbar$ only occur 
in class 4.
\item The `1-resolved' processes $\gamma\q \to \q\g$ and 
$\gamma\g \to \q\qbar$ occur in classes 2 and 5.
\item The `2-resolved' processes $\q\q' \to \q\q'$ (where $\q'$ 
may also represent an antiquark), $\q\qbar \to \q'\qbar'$,
$\q\qbar \to \g\g$, $\q\g \to \q\g$, $\g\g \to \q\qbar$ and
$\g\g \to \g\g$ occur in classes 1, 3 and 6.
\item Elastic, diffractive and low-$\pT$ events occur in class 1. 
\end{Itemize}
In the list above we have only indicated the lowest-order processes.
Within the context of the leading-log approximation, at least,
the subdivision into six event classes is easily generalized to
graphs with an arbitrary number of partons in the final state.
This classification is illustrated in Fig.~\ref{fig3} for a generic
ladder graph. To this picture final-state radiation can be added 
trivially. 

The notation direct, 1-resolved and 2-resolved is the conventional
subdivision of $\gamma\gamma$ interactions. The rest
is then called `soft-VMD'. As for the $\gamma\p$
case, our subdivision is an attempt to be more precise and internally
consistent than the conventional classes allow. One aspect is that
we really want to have a VMD$\times$VMD class that is nothing but
a scaled-down copy of the $\rho^0\rho^0$ and other vector-meson
processes, with a consistent transition between low-$\pT$ and
high-$\pT$ events (see below). Another aspect is that, in a 
complete description, the VMD and anomalous parts of the photon give
rise to different beam remnant structures, as discussed above, even 
when the hard subprocess itself may be the same.

A third aspect is that our subdivision provides further constraints; 
these, at least in principle, make the model more predictive. In
particular, the parton distributions of the photon are constrained
by the ansatz in eq.~(\ref{gammawavefunction}) to be given by 
\begin{equation}
f_a^{\gamma}(x,\mu^2) = f_a^{\gamma,\mrm{dir}}(x,\mu^2)
+ f_a^{\gamma,\mrm{VMD}}(x,\mu^2) 
+ f_a^{\gamma,\mrm{anom}}(x,\mu^2;k_0^2) ~.
\label{gammaPDF}
\end{equation}
Here
\begin{equation}
f_a^{\gamma,\mrm{dir}}(x,\mu^2) = Z_3 \, \delta_{a\gamma} \, 
\delta (1-x) 
\label{dirgammaPDF}
\end{equation}
and
\begin{equation}
f_a^{\gamma,\mrm{VMD}}(x,\mu^2) 
= \sum_{V = \rho^0,\omega,\phi,\Jpsi} 
\frac{4\pi\alpha}{f_V^2} f_a^{V}(x,\mu^2) ~.
\label{VMDgammaPDF}
\end{equation}
The anomalous part, finally, is fully calculable perturbatively, given
the boundary condition that the distributions should vanish for
$\mu^2 = k_0^2$. In principle, everything is therefore given.
In practice, the vector-meson distributions are not known, and so 
one is obliged to pick some reasonable ansatz with parameters fitted 
to the data. This is the approach taken in our SaS parameterizations
\cite{ourpdf}. By comparison, conventional distributions are defined 
for resolved processes only:
\begin{equation}
f_a^{\gamma,\mrm{res}}(x,\mu^2)  =  f_a^{\gamma,\mrm{VMD}}(x,\mu^2) +
f_a^{\gamma,\mrm{anom}}(x,\mu^2;k_0^2) ~.
\label{resgammaPDF}
\end{equation}
These resolved distributions are then less constrained, in particular  
with respect to the momentum sum \cite{momsum} of resolved partons.

\section{Cross Sections}

\subsection{The total cross section and its subdivision}

Total hadronic cross sections show a characteristic fall-off at 
low energies and a slow rise at higher energies. This behaviour 
can be parameterized by the form 
\begin{equation}
\sigma_{\mrm{tot}}^{AB}(s) = X^{AB} s^{\epsilon} + Y^{AB} s^{-\eta}
\label{sigmatotAB}
\end{equation}
for $A + B \to X$. The powers $\epsilon$ and $\eta$
are universal, with fit values \cite{DL92}
\begin{equation}
  \epsilon \approx 0.0808 ~, \qquad 
  \eta \approx 0.4525 ~,
\label{epsivalue}
\end{equation} 
while the coefficients $X^{AB}$ and $Y^{AB}$ are
process-dependent. Equation (\ref{sigmatotAB}) can be interpreted 
within Regge theory, where 
the first term corresponds to pomeron exchange and
gives the asymptotic rise of the cross section. Ultimately,
this increase violates the Froissart--Martin bound \cite{Froissart}; 
$\epsilon$ should therefore be thought of as slowly decreasing with 
energy (owing to multi-pomeron exchange effects), although data at 
current energies are well fitted by a constant $\epsilon$.
The second term, the reggeon one, is mainly of interest at low 
energies. For the purpose of our study we do not rely on the Regge 
interpretation of eq.~(\ref{sigmatotAB}), but can merely consider it as 
a convenient parameterization.

The VMD part of the $\gamma\p$ cross section should have a similar 
behaviour. The direct part reflects the parton distributions of the
proton; a small-$x$ behaviour like $xf(x) \sim x^{-\epsilon}$ would
give $\sigma_{\rm{dir}}^{\gamma\p} \sim s^{\epsilon}$.
The anomalous part is less easily classified: a purely perturbative
description would not give a behaviour like VMD, but a duality
argument with anomalous states interpreted in terms of higher 
vector-meson states would. Empirically, the $\gamma\p$
data are well described by 
\begin{equation}
\sigma_{\mrm{tot}}^{\gamma\p}(s) \approx 67.7 \, s^{\epsilon} + 
129 \, s^{-\eta} ~~[\mu\mrm{b}],
\label{sigtotgap}
\end{equation}
with $s$ in GeV$^2$. (Cross-sections are throughout given in
mb for hadron--hadron interactions, in $\mu$b for $\gamma$--hadron 
ones and in nb for $\gamma\gamma$ ones.) Actually, the above formula
is a prediction \cite{DL92} preceding the HERA data 
\cite{HERAtot,HERAtotnew}. 
The conclusion would seem to be that, at least as far as total
cross sections are concerned, the extended VMD description of
anomalous interactions is a reasonable first approximation.

If we then take the Regge-theory ansatz seriously also for the photon,
it is possible to derive an expression for the total $\gamma\gamma$ 
cross section
\begin{equation}
\sigma_{\mrm{tot}}^{\gamma\gamma}(s) \approx 211 \, s^{\epsilon} + 
215 \, s^{-\eta} ~~[\mrm{nb}].
\label{sigtotgaga}
\end{equation}
This is based on the assumption
that the pomeron and reggeon terms factorize,
$X^{AB} = \beta_{A\pom} \beta_{B\pom}$ and 
$Y^{AB} = \gamma_{A\reg} \gamma_{B\reg}$, so that
$X^{\gamma\gamma} = (X^{\gamma\p})^2/X^{\p\p}$ and
$Y^{\gamma\gamma} = 2(Y^{\gamma\p})^2/(Y^{\p\p}+Y^{\p\pbar})$,
with $X^{\p\p} \approx 21.70$ and 
$(Y^{\p\p}+Y^{\p\pbar})/2 \approx 77.23$.
In hadronic cross sections, factorization seems valid for the pomeron
term but not for the reggeon one, e.g. $X^{\p\pbar} = X^{\p\p}$ while
$Y^{\p\pbar} \approx 98.39 \gg Y^{\p\p} \approx 56.08$. The choice of
using the average of $Y^{\p\p}$ and $Y^{\p\pbar}$ then is an arbitrary one,
though it can be motivated roughly by arguments of counting the number
of allowed valence quark/antiquark annihilation/exchange diagrams 
possible in the various processes. The band of uncertainty can be 
obtained by using either $Y^{\p\p}$ or $Y^{\p\pbar}$ alone, i.e. 
$Y^{\gamma\gamma} = 297$ and $169$. This ambiguity only affects the 
low-energy behaviour, and so is not critical for us. It is illustrated
in Fig.~\ref{fig4}, where we also compare with existing data on
$\sigma_{\mrm{tot}}^{\gamma\gamma}(s)$.

Note that factorization is assumed to hold separately for the pomeron 
and the reggeon terms, not for the total cross section itself. That is, 
the relation $\sigma_{\mrm{tot}}^{\gamma\gamma} = 
2 (\sigma_{\mrm{tot}}^{\gamma\p})^2/(\sigma_{\mrm{tot}}^{\p\p}+
\sigma_{\mrm{tot}}^{\p\pbar})$
is not exact in this approach, although numerically it is a very
good approximation (usually to better than 1\%). 

Our eq.~(\ref{sigtotgaga}) above should be compared with 
the time-honoured expression $\sigma^{\gamma\gamma} = 240 + 270/W$
\cite{Rosner}. This corresponds to a critical pomeron, $\epsilon = 0$,
as was commonly assumed in the early seventies, and an $\eta = 0.5$,
but it is otherwise in the same spirit as our formula. Also numerically 
the two closely agree at not too large energies, see Fig.~\ref{fig4}. 

One should remember that our expression (\ref{sigtotgaga}) is here 
`derived' based on a simple Regge-theory ansatz that has no real
validity for the photon. Next we will proceed to study the 
contributions of the individual event classes. The constraints that come
from $\gamma\p$ physics data then directly feed into constraints on
the contribution from these classes and therefore on the total 
$\gamma\gamma$ cross section. At the end of the day we will therefore 
show that a cross section behaving roughly like 
eq.~(\ref{sigtotgaga}) should be a good approximation. In doing so, 
the properties of the event classes are also fixed, to a large extent.

Based on the subdivision into event classes, the total $\gamma\p$ cross
section may be written as
\begin{equation}
\sigma_{\mrm{tot}}^{\gamma\p} = \sigma_{\mrm{VMD}}^{\gamma\p} +
\sigma_{\mrm{dir}}^{\gamma\p} + \sigma_{\mrm{anom}}^{\gamma\p}
\label{sigdividegp}
\end{equation}
and the total $\gamma\gamma$ one as
\begin{equation}
\sigma_{\mrm{tot}}^{\gamma\gamma} =   
\sigma_{\mrm{VMD}\times\mrm{VMD}}^{\gamma\gamma} +
2 \sigma_{\mrm{VMD}\times\mrm{dir}}^{\gamma\gamma} +
2 \sigma_{\mrm{VMD}\times\mrm{anom}}^{\gamma\gamma} +
\sigma_{\mrm{dir}\times\mrm{dir}}^{\gamma\gamma} +
2 \sigma_{\mrm{dir}\times\mrm{anom}}^{\gamma\gamma} +
\sigma_{\mrm{anom}\times\mrm{anom}}^{\gamma\gamma} ~.
\label{sigdividegg}
\end{equation}
Here we explicitly keep the factor of 2 for the off-diagonal terms,
where the r\^ole of the two incoming photons may be interchanged. 

\subsection{The VMD contributions}

The $V\p$ cross sections may be parameterized as
\begin{eqnarray}
\sigma_{\mrm{tot}}^{\rho^0\p} \approx \sigma_{\mrm{tot}}^{\omega\p} 
  & \approx & \frac{1}{2} \left( \sigma_{\mrm{tot}}^{\pi^+\p} +
  \sigma_{\mrm{tot}}^{\pi^-\p} \right) 
  \approx 13.63 \, s^{\epsilon} + 31.79 \, s^{-\eta} ~~[\mrm{mb}],
\label{sigrho} \\
\sigma_{\mrm{tot}}^{\phi\p} & \approx & 
  \sigma_{\mrm{tot}}^{\K^+\p} + \sigma_{\mrm{tot}}^{\K^-\p} - 
  \sigma_{\mrm{tot}}^{\pi^-\p} 
  \approx 10.01 \, s^{\epsilon} - 1.51 \, s^{-\eta} ~~[\mrm{mb}].
\label{sigphi}
\end{eqnarray}
The $\phi\p$ cross section is not expected to have a reggeon term
and indeed the additive quark model \cite{addquark} formulae give a 
contribution close to zero; a small negative term could easily come 
from threshold effects and so we choose to keep it. 
Lacking measurements of $\D\p$ cross sections we cannot use the
additive quark model to estimate the $\Jpsi\p$ cross section. The latter 
could in principle be extracted from data on $\Jpsi$ production in 
nuclear collisions. Superficially a value of about 6~mb at 
$5 \lessim \sqrt{s} \lessim 10\,$GeV comes out but this value 
presumably is too large since in the so far accessible kinematical range
of large positive $x_F$ the $\Jpsi$ is formed outside the nucleus. 
Therefore we fix the VMD coupling of the $\Jpsi$ at its leptonic 
value and use a low-energy measurement of elastic $\Jpsi$ photoproduction 
to determine the $\Jpsi\p$ cross section. This implies a reduction 
of the $\psi\p$ cross section by a factor of about 10 compared to
the $\phi\p$ one, in agreement with the expectation that the soft pomeron
couples more weakly to heavier quarks. 
Again using factorization for the pomeron and reggeon terms 
separately, the total cross section for two vector mesons is
\begin{equation}
\sigma_{\mrm{tot}}^{V_1 V_2} \approx 
\frac{X^{\p V_1} X^{\p V_2}}{X^{\p\p}} \, s^{\epsilon} +
\frac{2Y^{\p V_1} Y^{\p V_2}}{Y^{\p\p} + Y^{\p\pbar}} \, s^{-\eta} ~.
\end{equation}
These $X$ and $Y$ coefficients are collected in Table~\ref{tab1}. 

The total VMD cross sections are obtained as weighted sums of
the allowed vector-meson states,
\begin{eqnarray}
\sigma_{\mrm{VMD}}^{\gamma\p} & = & 
\sum_V \frac{4\pi\alphaem}{f_V^2} \, \sigma_{\mrm{tot}}^{V\p} 
\approx 54 \, s^{\epsilon} + 115 \, s^{-\eta} ~~[\mu\mrm{b}],
\label{sigVMDgap} \\
\sigma_{\mrm{VMD}\times\mrm{VMD}}^{\gamma\gamma} & = &
\sum_{V_1} \frac{4\pi\alphaem}{f_{V_1}^2}
\sum_{V_2} \frac{4\pi\alphaem}{f_{V_2}^2} \,
\sigma_{\mrm{tot}}^{V_1 V_2} 
\approx 133 \, s^{\epsilon} + 170 \, s^{-\eta} ~~[\mrm{nb}].
\label{sigVMDgaga}
\end{eqnarray}
In Fig.~\ref{fig5} we show the breakdown of 
$\sigma_{\mrm{VMD}\times\mrm{VMD}}^{\gamma\gamma}$
by vector-meson combination. Obviously the $\rho^0\rho^0$
combination dominates. 

For a description of VMD events, a further subdivision into elastic 
(el), diffractive (sd and dd for single and double diffractive) 
and non-diffractive (nd) events is required. Keeping only the 
simplest diffractive topologies, one may write
\begin{equation}
\sigma_{\mrm{tot}}^{AB}(s) = \sigma_{\mrm{el}}^{AB}(s) +
\sigma_{\mrm{sd}(XB)}^{AB}(s) + \sigma_{\mrm{sd}(AX)}^{AB}(s) +
\sigma_{\mrm{dd}}^{AB}(s) + \sigma_{\mrm{nd}}^{AB}(s)~.
\end{equation}
The elastic and diffractive cross sections for all required $V\p$ and
$V_1 V_2$ processes have been calculated and parameterized in the
context of our model presented in ref.~\cite{haha}. The same formulae
are used as those collected in section 4 of that paper, and so are
not repeated here; only the expressions in its eq.~(26) have to be 
replaced. The following parameterizations have been chosen:
\begin{eqnarray}
M_{\mrm{max},XB}^2 & = & c_1 s + c_2 ~, \nonumber \\
B_{XB} & = & c_3 + \frac{c_4}{s} ~, \nonumber \\
M_{\mrm{max},AX}^2 & = & c_5 s + c_6 ~, \nonumber \\
B_{AX} & = & c_7 + \frac{c_8}{s} ~, \nonumber \\
\Delta_0 & = & d_1 + \frac{d_2}{\ln s} + 
\frac{d_3}{\ln^2 s} ~, \nonumber \\
M_{\mrm{max},XB}^2 & = & s \left( d_4 + \frac{d_5}{\ln s} + 
\frac{d_6}{\ln^2 s} \right) ~, \nonumber \\
B_{XX} & = & d_7 + \frac{d_8}{\sqrt{s}} + \frac{d_9}{s}~.
\label{diffpara}  
\end{eqnarray}
The coefficients $c_i$ and $d_i$ are given in Table~\ref{tab1}.
Additionally the $b$ slope parameters are $b_{\p} = 2.3$~GeV$^{-2}$,
$b_{\rho} = b_{\omega} = b_{\phi} = 1.4$~GeV$^{-2}$ and
$b_{\Jpsi} = 0.23$~GeV$^{-2}$. 
The non-diffractive cross-section is then given by whatever is left.
This subdivision is shown in Fig.~\ref{fig6}
for the sum of all meson combinations, which then mainly reflects
the $\rho^0\rho^0$ composition.

The $\sigma_{\mrm{nd}}$ may be further subdivided into a low-$\pT$ 
and a high-$\pT$ class. Since the $2\to2$ parton--parton scattering 
cross sections are divergent in the limit $\pT \to 0$, some further 
care is needed for this classification. We expect the perturbative 
formulae to break down at small $\pT$, since an exchanged gluon 
with a large transverse wavelength $\lambda_{\perp} \sim 1 / \pT$
cannot resolve the individual colour charges inside a hadron. 
The hadron being a net colour singlet, the effective 
coupling should therefore vanish in this limit. A parameter 
$\pTmin = \pTmin(s)$ is introduced to describe the border
down to which the perturbative expression is assumed to be valid
\cite{gammap}. The jet rate above $\pTmin$ may still be large, in 
fact even larger than the total $\sigma_{\mrm{nd}}$. It is therefore
necessary to allow for the possibility of having several perturbative
parton--parton interactions in one and the same event, i.e. to 
unitarize the jet emission probability. We do this using the 
formalism of ref. \cite{TSMZ}. A fit to collider multiplicities 
gives 
\begin{equation}
\pTmin(s) = \pTmin^{\mrm{VMD}}(s) \approx 1.30 + 0.15 \, 
\frac{\ln(E_{\mrm{cm}}/200)}{\ln(900/200)} ~~[\mrm{GeV}]~.  
\label{pTmins}
\end{equation}
Here we uses the CTEQ 2L \cite{protonpdf} leading-order 
parton distributions (extended to small $x$ and $Q^2$ as described in
\cite{gammap}), with $\pT^2$ as scale choice. 

\subsection{The direct and anomalous contributions}
\label{subsectdiranom}

Comparing eqs. (\ref{sigtotgap}) and (\ref{sigVMDgap}), about 80\%
of the $\gamma\p$ total cross section is seen to come from the VMD
term. The remaining 20\% is to be attributed to the direct and anomalous
components. When applying a perturbative description, the anomalous part 
is negligible at small energies. The dependence of the direct cross 
section on $k_0$ can then be used to determine this parameter. We obtain 
a value of $k_0 \approx 0.5$~GeV \cite{gammap}, which is consistent with 
the simple-minded answer $k_0 \approx m_{\phi}/2$. In our study of the
parton distributions of the photons \cite{ourpdf} a reasonable 
$f_a^{\gamma,\mrm{res}}(x,\mu^2)$ was obtained with $Q_0 = 0.6$~GeV,
i.e. the same order. For this study we have stayed with the latter 
number, $k_0 = 0.6$~GeV.

The anomalous process contains two cut-off parameters, the $k_0$ scale
for the photon to branch to a perturbative $\q\qbar$ pair and a
$\pTmin^{\mrm{anom}}$ scale for one of the anomalous-photon partons 
to interact in a hard process. As a first guess, one might choose
$\pTmin^{\mrm{anom}}$ also to be given by eq.~(\ref{pTmins}).
However, this turns out to give a cross section increasing too rapidly
at large energies. 
Physically, it is understandable why hard processes should be
more suppressed at small $\pT$ in anomalous processes than in VMD ones: 
the anomalous photon corresponds to a $\q\qbar$ pair of larger virtuality 
than a VMD one, and hence of smaller spatial extent, i.e. with
larger potential for colour screening. The best recipe for
including this physics aspect is not well understood. As a purely
pragmatical recipe, one can pick $\pTmin^{\mrm{anom}}(s)$ with an $s$
dependence such that the VMD, direct and anomalous processes add up 
to the expected behaviour (\ref{sigtotgap}). Over the energy range
$20 \lessim \sqrt{s} \lessim 1000$ a suitable parameterization then is
\begin{equation}
\pTmin^{\mrm{anom}}(s) \approx 0.6 + 0.125 \, \ln^2(1 + \sqrt{s}/10) 
~~[\mrm{GeV}] ~.
\label{pTminanom}
\end{equation}
This is based on parton distributions SaS 1D \cite{ourpdf} for the 
photon and CTEQ 2L 
for the proton, combined with lowest-order matrix elements. At low
energies the results are fairly unstable to variations in $k_0$,
and so the behaviour of $\pTmin^{\mrm{anom}}(s)$ in this region should 
not be over-interpreted. The split of the $\gamma\p$ cross section into 
VMD, direct and anomalous is shown in Fig.~\ref{fig7}. 

In this purely perturbative picture we have neither included any class 
of `soft' anomalous interactions nor the possibility of multiple
parton--parton interactions. Keeping everything  else the same, the 
former would increase the cross section and the latter decrease it.
However, when introducing a soft component, it is important to avoid
double-counting. To illustrate the issue, consider the simple graph of 
Fig.~\ref{fig8}a. There are two transverse momentum scales, $\kT$ and 
$\pT$. It is a simpler version of Fig.~\ref{fig3}, with inessential
gluons removed and for $\gamma\p$ rather than $\gamma\gamma$, so has
one scale less. The allowed phase space can then conveniently be 
represented by a two-dimensional plane, Fig.~\ref{fig8}b. The region 
$\kT < k_0$ corresponds to a small transverse momentum at the 
$\gamma \to \q\qbar$ vertex, and thus to VMD processes. For $\kT > k_0$, 
the events are split along the diagonal $\kT = \pT$. If $\kT > \pT$, 
the hard $2 \to 2$ process of Fig.~\ref{fig8}a is $\gamma\g \to \q\qbar$, 
and the lower part of the graph is part of the leading log QCD evolution 
of the gluon distribution inside the proton. These events are direct 
ones. If $\pT > \kT$, on the other hand, the hard process is 
$\qbar \q' \to \qbar \q'$, and the $\gamma \to \q\qbar$ vertex builds 
up the quark distribution inside a photon. These events are thus 
anomalous ones.

By analogy with the VMD representation, each fixed-$\kT$ component of
$\gamma \leftrightarrow \q\qbar$ fluctuations can be considered as
a separate `hadron species', with a density of states proportional to
$\d \kT^2/\kT^2$. Each vertical `tower' at some given $\kT$ scale would 
correspond to a higher excited vector resonance in the context of
a generalized VMD model. In this
tower, the soft events would be in the direct sector and the hard 
events in the anomalous sector. In the region of large $\kT$ values 
the perturbative language is well defined, and no problems should arise. 
As smaller and smaller $\kT$'s are considered, however, one could expect 
event properties that are intermediate to those of VMD. In particular, 
multiple parton--parton interactions could be possible, and this would 
affect the relation between calculated jet cross sections and the total
event cross section. Previously we had to introduce a large 
$\pTmin^{\mrm{anom}}$ scale at high energies to solve the problem of 
too large an anomalous cross section, which means we left an 
un-populated hole in the middle of Fig.~\ref{fig8}b (indicated by a 
question mark). The hope is that multiple interactions would provide a 
natural resolution of this problem, in the sense that most anomalous 
events have one hard scattering above $\pTmin^{\mrm{anom}}$, while the 
anomalous region with $\pT < \pTmin^{\mrm{anom}}$ does not significantly 
contribute new events but rather further interactions inside the events 
above. 

The hope for a simple solution is not borne out by studies, however 
\cite{previoustwo}. Eikonalization does dampen the increase of the
anomalous and direct cross sections, but not enough. With a 
$\pTmin^{\mrm{anom}}(s) = \pTmin^{\mrm{VMD}}(s)$ the cross section is 
still increasing too fast at high energies, if the pomeron-style
behaviour is taken as reference. Problems that appear already in
the $\gamma\p$ sector are even more severe in attempts at a
$\gamma\gamma$ description along the same lines. It therefore seems 
clear that further aspects have to be taken into account, such as 
momentum conservation, coherence effects or strict geometrical cuts.
 
We intend to return to these problems, but for the moment stay with a
purely perturbative description of the direct and anomalous components.
By pushing this approach to its logical conclusion, we see what to
expect from it and what limitations it has. Furthermore, for most
event properties we expect the perturbative description to be 
perfectly adequate. 

\subsection{The total $\gamma\gamma$ cross section by component}

Turning to the $\gamma\gamma$ cross sections, in principle all
free parameters have now been fixed, and the cross section for each of
the six event classes can be obtained. The VMD$\times$VMD one has 
already been discussed; the others are given as integrals of $2 \to 2$ 
scattering cross sections above the respective $\pT$ cut-offs already
specified. The results are shown in Fig.~\ref{fig9}, class by class. 
For comparison, we also show the results that would be obtained in
a simple factorization ansatz
\begin{equation}
\sigma_{i \times j}^{\gamma\gamma} =
\frac{2}{1+\delta_{ij}} \,
\frac{2 \, \sigma_i^{\gamma\p} \, \sigma_j^{\gamma\p}}%
{\sigma_{\mrm{tot}}^{\p\p} + \sigma_{\mrm{tot}}^{\p\pbar}} ~,
\label{factorizedtot}
\end{equation}
where $i,j=$ VMD, direct and anomalous.

A few comments about each of the classes: 
\begin{Enumerate}
\item For the VMD$\times$VMD class in principle the simple 
factorization ansatz is exact in our model; some minor deviations
come from the reggeon term.
\item The VMD$\times$direct class comes out about a factor $3/2$ 
larger than expected from the factorized ansatz. This difference
can be understood by comparing the jet and the total cross sections
of a proton and a $\pi$ meson, where the latter is taken as
a prototype for a VMD meson. Both the $\p$ and the $\pi$ have
parton distributions normalized to unit momentum sum, and the
same small-$x$ behaviour (using our prescription \cite{gammap}).
Neglecting some differences in the shape of the parton 
distributions, the jet rates therefore are comparable between
$\p\p$, $\pi\p$ and $\pi\pi$ collisions. The total cross sections,
on the other hand, are scaled down roughly by a factor $2/3$
between $\p\p$ and $\pi\p$. Therefore the jet rate per event
is a factor $3/2$ larger for $\pi\p$ than for $\p\p$, and it is 
this factor that appears above. That is, eq.~(\ref{factorizedtot})
would have worked only if $\gamma\pi/\pi\pi$ cross sections 
could have been used rather than $\gamma\p/\p\p$ ones. The larger
jet rate per event for mesons should be reflected in differences 
in the eikonalization treatment of the direct and anomalous 
components of $\gamma\p$ and $\gamma\pi$ events.
\item The VMD$\times$anomalous component gives exactly the same 
factor $3/2$ mismatch as discussed above for the VMD$\times$direct
one.
\item The direct$\times$direct component is not at all well 
predicted by the factorized ansatz. The latter yields a cross 
section growing at large energies at a rate related to the 
small-$x$ behaviour of the proton distribution functions, i.e.\ 
$\propto s^{\epsilon}$ for our modified distributions. On the 
other hand, the total cross section for 
$\gamma\gamma \to \q\qbar$ is proportional to $\ln(s/k_0^2)/s$, 
and thus drops rapidly with c.m. energy. 
\item The direct$\times$anomalous component compares reasonably
well with the prediction from factorization.
\item The anomalous$\times$anomalous process, finally, is most 
uncertain, since it completely involves the interactions of the
least well understood component of the photon wave function.
\end{Enumerate}

In Fig.~\ref{fig10} the total $\gamma\gamma$ cross section is
compared between the Regge type ansatz~(\ref{sigtotgaga}) and
the sum of the six classes above, eq.~(\ref{sigdividegg}).
It should be remembered that the first three are the 
dominant ones. In fact, since the direct and anomalous components 
together give about 20\% of the
$\gamma\p$ total cross section, the expectation is that the last
three classes together would only give a 4\% contribution to the 
total $\gamma\gamma$ cross section. The anomalous$\times$anomalous 
component may give a somewhat larger contribution than expected, but 
still the 4\% number gives the right ballpark.
The first three classes, on the other hand, are all 
related to the respective $\gamma\p$ classes, with only a replacement
of a $\p$ by a $V$ (and an extra weight factor $4\pi\alphaem/f_V^2$).
Apart from the appearance of a factor $3/2$ in the VMD$\times$direct 
and VMD$\times$anomalous components, which should (largely if not
completely) go away in a fully
eikonalized description, these components behave as expected. 
This makes the argumentation for eq.~(\ref{sigtotgaga}) credible.
However, if one wants to take a conservative approach, the 
spread between the two curves in Fig.~\ref{fig10} could be viewed
as a band of uncertainty. The data are not yet precise to provide
any discrimination, cf. Fig.~\ref{fig4}.

One can also compare our $\sigma_{\mrm{tot}}^{\gamma\gamma}$
with the numbers obtained in various minijet-based approaches 
\cite{minijet,LEP2workshop}. For $E_{\mrm{cm}} = 200$~GeV, cross sections 
in the range 1000--1800 nb are typically obtained, but are reduced to
about 500 nb if unitarity is enforced, in agreement with our results.

\section{Event Properties}

The subdivision of the total $\gamma\p$ and $\gamma\gamma$ cross 
sections above, with the related choices of cut-off parameters etc.,
specifies the event composition at the hard-scattering level. 
Some interesting observables can be based on this classification
alone. For instance, Fig~\ref{fig11} gives the cross section for
elastic events of the kinds $\rho^0\rho^0$, $\rho^0\omega$,
$\rho^0\phi$ and $\rho^0\Jpsi$. The former three processes are good tests 
for the validity of the VMD ansatz, whereas the last one could provide 
new insights.

For most studies it is necessary to consider
the complete event structure, i.e. to add models 
for initial- and final-state QCD radiation (parton showers), for beam
remnants, and for fragmentation and secondary decays \cite{gammap}.
A Monte Carlo generation of complete hadronic final states is obtained
with {\sc Pythia}/{\sc Jetset} \cite{PyJe}. 
Thus any experimental quantity can be studied. This section gives some 
representative examples. In particular, we compare the properties
of $\p\p$, $\gamma\p$ and $\gamma\gamma$ events. It should be noted
that $\p\p$ and $\p\pbar$ events are very similar for the 
quantities studied here. Unless otherwise specified, the figures 
refer to an $E_{\mrm{cm}} = \sqrt{s_{\gamma\gamma}} = 200$~GeV. As we 
will show at the end of the section, the qualitative features do not 
depend critically on this choice. Furthermore, figures relevant for 
LEP~2 energies can be found in the proceedings of the LEP~2 workshop
\cite{LEP2workshop,LEP2generator}, so it makes sense to complement 
here with the higher-energy behaviour relevant for future linear 
colliders.

Figure~\ref{fig12} shows the total transverse energy per event for each
of the six components of the $\gamma\gamma$ cross section. The spike
at small $\sum E_{\perp}$ for the VMD$\times$VMD class comes from 
elastic scattering events, e.g. $\gamma\gamma \to \rho^0\rho^0$.
Also diffractive events contribute in this region. The 
large-$\sum E_{\perp}$ tail of the VMD$\times$VMD curve is enhanced
by the possibility of multiple parton--parton interactions, which is
only included for this class currently. Because of the 
larger $\pTmin^{\mrm{anom}}$ cut-off, the classes involving anomalous
photons typically have larger $\sum E_{\perp}$, while the smaller
$p_0$ cut-off for the direct processes corresponds
to smaller median $\sum E_{\perp}$. However, note that 
the $\gamma\gamma \to \q\qbar$ processes only fall off very slowly with
$\pT$, in part because of the absence of structure functions, in part
because of the form of the matrix element itself. The 
direct$\times$direct class therefore wins out at very large  
$\sum E_{\perp}$. 

The results of Fig.~\ref{fig12} are a bit misleading, since the relative 
importance of the six event classes is not visible. The weighted 
mixture is shown in Fig.~\ref{fig13}, also compared with $\gamma\p$ and 
$\p\p$ events. One observes a steady progression, with
$\langle \sum E_{\perp} \rangle_{\p\p} < 
\langle \sum E_{\perp} \rangle_{\gamma\p} < 
\langle \sum E_{\perp} \rangle_{\gamma\gamma}$.
This pattern, of more activity for a $\gamma$ than for a $\p$,
is seen in essentially all distributions. The elastic spike at small
$\sum E_{\perp}$ is less pronounced for $\gamma\gamma$, owing to three
factors: the VMD$\times$VMD component is only a part of the 
$\gamma\gamma$ cross section, elastic scattering is a smaller fraction 
of the total $\rho^0\rho^0$ cross section than it is for $\p\p$, and 
kinetic energy in the $\rho^0 \to \pi^+\pi^-$ decays add to the total 
transverse energy. 

The $E_{\perp}$ flow as a function of pseudorapidity, 
$\d E_{\perp} / \d \eta$, is given in Fig.~\ref{fig14}. 
It illustrates how $\gamma\p$ interpolates between $\p\p$ and 
$\gamma\gamma$: around the direction of the incoming photon, the 
$\gamma\p$ events look like the $\gamma\gamma$ ones, while they look
more like $\p\p$ ones in the opposite direction, with an intermediate
behaviour in the central region. Since elastic and single diffractive 
events are likely to be removed from `minimum-bias' data samples,  
Fig.~\ref{fig15} shows the behaviour without those two event classes.
The quantitative $\gamma\gamma/\gamma\p/\p\p$ differences then are
slightly reduced, but qualitatively remain unchanged.
Also in subsequent figures these events have been removed, and the
same comment can be made there. 

The charged-multiplicity distributions follow essentially the same
pattern as shown for the $\sum E_{\perp}$ ones in 
Figs.~\ref{fig12}--\ref{fig15},
and are therefore not included here. There is one noteworthy
exception, however: the direct$\times$direct component does not have
a tail out to large multiplicities. That is, even if the process 
$\gamma\gamma \to \q\qbar$ can generate large $\pT$ values, the
absence of any beam jets keeps the multiplicity down. 

The transverse momentum spectrum of charged particles is shown in 
Fig.~\ref{fig16}. The larger high-$\pT$ tail of the $\gamma\gamma$ 
processes is one of the simplest observables to experimentally 
establish differences between $\p\p$, $\gamma\p$ and $\gamma\gamma$. 
Of course, the cause of the differences is to be sought in the higher
jet rates associated with photon interactions. The jet spectra are
compared in Fig.~\ref{fig17}, using a simple cone algorithm where 
a minimum $E_{\perp}$ of 5 GeV is required inside a cone of 
$\Delta R = \sqrt{ (\Delta\eta)^2 + (\Delta\phi)^2} < 1$. Already 
for an $E_{\perp\mrm{jet}}$ of 5 GeV there are more than three times 
as many jets in  $\gamma\gamma$ as in $\p\p$, and this ratio then 
increases with increasing $E_{\perp\mrm{jet}}$. 
The spectacular differences in the jet rate at large $\pT$
are highlighted in Fig.~\ref{fig18}. They mainly come about because 
the direct component involves the full energy of the incoming photon.
The pseudorapidity distribution of jets is shown in Fig.~\ref{fig19}.
As in the inclusive $\d E_{\perp}/\d\eta$ distributions, the difference
in behaviour between the $\gamma$ and $\p$ hemispheres is readily 
visible.

To illustrate the energy dependence of these distributions, 
Fig.~\ref{fig20} gives the $\d E_{\perp} / \d\eta$ flow for 
c.m. energies of 50 GeV. This can be compared with the
result for 200 GeV in Fig.~\ref{fig15}. Qualitatively, the same pattern 
is seen at both energies, although relative differences tend to be 
somewhat reduced at larger energies. This is also true for other
observables, such as jet rates. One reason is that the possibility of 
multiple parton--parton interactions in the VMD component pushes 
up the activity in those events at larger energies, and thus brings 
them closer to the anomalous class. The importance of the direct class, 
on the other hand, is reduced at large energies. Further, at large 
energies, jet production is dominantly initiated by small-$x$ 
incoming partons, where the VMD and anomalous distributions are 
more similar than at large $x$ (although still different).

\section{Summary and Outlook}

In this paper we have shown that our model for $\gamma\p$ events 
\cite{gammap} can be consistently generalized to $\gamma\gamma$ events.
That is, essentially all free parameters are fixed by 
(low-energy) $\gamma\p$
phenomenology. Since we start out with a more detailed subdivision
of the $\gamma\p$ total cross section than has conventionally been 
done in the past, our $\gamma\gamma$ model also contains a
richer spectrum of possible processes. We distinguish six main event
classes, but most of these contain further subdivisions. The aim is
that this approach will allow predictions for a broader range of
observables than is addressed in conventional models. 
For instance, although not discussed in detail here, our approach
does correlate the hard-jet physics in the central rapidity region 
with the structure of the beam remnants.

This does not mean that all results are complicated. We have shown
that the simple Regge-theory expression 
$\sigma_{\mrm{tot}}^{\gamma\gamma}(s) \approx 
211 \, s^{0.08} + 215 \, s^{-0.45}$~[nb] comes close to what is
obtained in our full analysis, and have a fair understanding 
where differences come from. We therefore expect this expression 
to be good to better than 10\% from a few GeV onwards, at least to 
the top $\gamma\gamma$ energies that could be addressed with the next 
generation of linear $\ee$ colliders. Also global event properties
show a very simple pattern, with more activity (transverse energy,
multiplicity, jets, \ldots) in $\gamma\p$ events than in $\p\p$ ones,
and still more in $\gamma\gamma$ ones. This is perhaps contrary to 
the na\"{\i}ve image of a `clean' point-like photon.
The $\gamma\p$ events show
their intermediate status by having a photon (proton) hemisphere that 
looks much like $\gamma\gamma$ ($\p\p$) events, with a smooth
interpolation in the middle.

In our current model the perturbative approach to the description of 
the direct and anomalous components is pushed to its extreme. In this 
sense it is a useful study. However, we also see that it has its 
limitations: at high energies a purely perturbative treatment of the 
direct and anomalous components is no longer possible and unitarity
corrections have to be taken into account, possibly through
an eikonalized treatment of the two components along the lines
indicated above (i.e.\ treating the direct event class as the
soft component of the anomalous one). The goal is a consistent
treatment covering the whole $(k_{\perp 1},k_{\perp 2},\pT)$ volume 
of $\gamma\gamma$ events in a consistent fashion, with smooth 
transitions between the various regions. Unfortunately this is a not 
so trivial task, and anyway must be based on input from the simpler 
model above. The current model therefore is a useful step towards 
an improved understanding of the photon and its interactions.
 
Also many other aspects need to be studied. Disagreements between
the HERA data and our model can be found for the profile of beam jets, 
the structure of underlying events, the event topology composition 
and so on, indicating the need for further refinements 
\cite{HERAwork}. The transition from a 
real $\gamma$ to a virtual $\gamma^*$ is still not well understood. 
Production in diffractive systems currently attract much attention.
Further issues could be mentioned, but the conclusion must be that
much work remains before we can claim to have a complete overview
of the physics involved in $\gamma\p$ and $\gamma\gamma$ events, 
let alone understand all the details.

\clearpage

\begin{table}[p]
\begin{tabular}{|c||c|c|c||c|c|c|c|c|c|}
\hline
   & $\rho^0\p$ & $\phi\p$ & $\Jpsi\p$ & $\rho^0\rho^0$ & $\phi\rho^0$
   & $\Jpsi\rho^0$ & $\phi\phi$ & $\Jpsi\phi$ & $\Jpsi\Jpsi$ \\ 
\hline \hline
$X$ & 13.63 & 10.01 & 0.970 & 8.56 & 6.29 & 0.609 & 4.62 & 
    0.447 & 0.0434 \\
$Y$ & 31.79 & $-1.51$ & $-0.146$ & 13.08 & $-0.62$ & $-0.060$ &
0.030 & $-0.0028$ & 0.00028 \\ 
\hline \hline
$c_1$ & 0.213   & 0.213   & 0.213   & 0.267   & 0.267 
      & 0.267   & 0.232   & 0.232   & 0.115   \\
$c_2$ & 0       & 0       & 7       & 0       & 0  
      & 6       & 0       & 6       & 5.5     \\ \hline
$c_3$ & $-0.47$ & $-0.47$ & $-0.55$ & $-0.46$ & $-0.48$ 
      & $-0.56$ & $-0.48$ & $-0.56$ & $-0.58$ \\
$c_4$ & 150     & 150     & 800     & 75      & 100
      & 420     & 110     & 470     & 570     \\ \hline
$c_5$ & 0.267   & 0.232   & 0.115   & 0.267   & 0.232
      & 0.115   & 0.232   & 0.115   & 0.115   \\
$c_6$ & 0       & 0       & 0       & 0       & 0  
      & 0       & 0       & 0       & 5.5     \\ \hline
$c_7$ & $-0.47$ & $-0.47$ & $-0.47$ & $-0.46$ & $-0.46$
      & $-0.50$ & $-0.48$ & $-0.52$ & $-0.58$ \\
$c_8$ & 100     & 110     & 110     & 75      & 85  
      & 90      & 110     & 120     & 570     \\ \hline
$d_1$ & 3.11    & 3.12    & 3.13    & 3.11    & 3.11 
      & 3.12    & 3.11    & 3.18    & 4.18    \\
$d_2$ & $-7.10$ & $-7.43$ & $-8.18$ & $-6.90$ & $-7.13$ 
      & $-7.90$ & $-7.39$ & $-8.95$ & $-29.2$ \\
$d_3$ & 10.6    & 9.21    & $-4.20$ & 11.4    & 10.0    
      & $-1.49$ & 8.22    & $-3.37$ & 56.2    \\ \hline
$d_4$ & 0.073   & 0.067   & 0.056   & 0.078   & 0.071
      & 0.054   & 0.065   & 0.057   & 0.074   \\
$d_5$ & $-0.41$ & $-0.44$ & $-0.71$ & $-0.40$ & $-0.41$
      & $-0.64$ & $-0.44$ & $-0.76$ & $-1.36$ \\
$d_6$ & 1.17    & 1.41    & 3.12    & 1.05    & 1.23 
      & 2.72    & 1.45    & 3.32    & 6.67    \\ \hline
$d_7$ & $-1.41$ & $-1.35$ & $-1.12$ & $-1.40$ & $-1.34$
      & $-1.13$ & $-1.36$ & $-1.12$ & $-1.14$ \\
$d_8$ & 31.6    & 36.5    & 55.2    & 28.4    & 33.1
      & 53.1    & 38.1    & 55.6    & 116.2   \\
$d_9$ & 95      & 132     & 1298    & 78      & 105
      & 995     & 148     & 1472    & 6532    \\
\hline
\end{tabular}
\caption%
{Coefficients of the total and partial $V\p$ and $V_1V_2$ cross 
sections, according to the formulae given in the text,
eqs.~(\ref{sigmatotAB}) and (\ref{diffpara}). The $\omega$
is not shown separately, since it is assumed to have the same 
behaviour as the $\rho^0$.
\label{tab1}}
\end{table}

\clearpage
      
\begin{figure}[p]
\begin{center}
\mbox{\epsfig{file=gagafig1.ps,bbllx=0,bblly=0,bburx=382,bbury=100}}
\end{center}
\vspace{2mm}
\caption%
{Contributions to hard $\gamma\p$ interactions: a) VMD, 
b) direct, and c)~anomalous. Only the basic graphs are illustrated;
additional partonic activity is allowed in all three processes.
The presence of spectator jets has been indicated by dashed lines,
while full lines show partons that (may) give rise to 
high-$\pT$ jets.
\label{fig1}}
\end{figure}

\vspace{2cm}
 
\begin{figure}[p]
\begin{center}
\mbox{\epsfig{file=gagafig2.ps,bbllx=0,bblly=0,bburx=382,bbury=230}}
\end{center}
\vspace{2mm}   
\caption%
{The six classes contributing to hard $\gamma\gamma$ interactions: 
a) VMD$\times$VMD, 
b)~VMD$\times$direct, c) VMD$\times$anomalous, 
d) direct$\times$direct, e) direct$\times$anomalous, 
and f)~anomalous$\times$anomalous. 
Notation as in Fig.~\protect\ref{fig1}.
\label{fig2}}
\end{figure}

\clearpage
   
\begin{figure}[p]
\begin{center}
\mbox{\epsfig{file=gagafig3.ps,bbllx=0,bblly=0,bburx=382,bbury=230}}
\end{center}
\vspace{-2mm}   
\caption%
{A generic Feynman diagram (in the leading-log approximation)
for $\gamma\gamma$ interactions and its decomposition into
six components.
\label{fig3}}
\end{figure}

\begin{figure}[p]
\begin{center}
\mbox{\epsfig{file=gagafig4.eps}} 
\end{center}
\vspace{-5mm}   
\caption%
{The total $\gamma\gamma$ cross section. Full curve: the 
parameterization of eq.~(\protect\ref{sigtotgaga}). 
Dashed curves: range obtained by varying $Y^{\gamma\gamma}$ as 
described in the text. Dashed-dotted curve: the critical-pomeron 
parameterization \protect\cite{Rosner}. 
Data points: open triangles PLUTO 1984, filled triangles PLUTO 1986,
squares TPC/2$\gamma$ 1985, spades TPC/2$\gamma$ 1991, circles MD-1 
1991, full square CELLO 1991 \protect\cite{MD1}.
\label{fig4}}
\end{figure}
 
\clearpage

\begin{figure}[p]
\begin{center}
\mbox{\epsfig{file=gagafig5.eps}}
\end{center}
\vspace{-5mm}   
\caption%
{The total VMD$\times$VMD cross section, full curve, and its subdivision
by vector-meson combination. The components are separated by dashed 
curves, from bottom to top: $\rho^0\rho^0$, $\rho^0\omega$, $\rho^0\phi$,
$\rho^0\Jpsi$, $\omega\omega$, $\omega\phi$, $\omega\Jpsi$, $\phi\phi$,
$\phi\Jpsi$, and $\Jpsi\Jpsi$. Some of the latter components are too 
small to be resolved in the figure. 
\label{fig5}}
\end{figure}

\begin{figure}[p]
\begin{center}
\mbox{\epsfig{file=gagafig6.eps}} 
\end{center}
\vspace{-5mm}   
\caption%
{The total VMD$\times$VMD cross section, full curve, and its subdivision
by event topology. The components are separated by dashed curves, from
bottom to top: elastic, single diffractive (split for the two sides by 
the dotted curve), double diffractive, and non-diffractive (including 
jet events unitarized).
\label{fig6}}
\end{figure}

\clearpage

\begin{figure}[p]
\begin{center}
\mbox{\epsfig{file=gagafig7.eps}} 
\end{center}
\vspace{-5mm}   
\caption%
{Components of the $\gamma\p$ cross section: lower dashed
curve the VMD contribution, middle dashed VMD$+$direct
and upper dashed VMD$+$direct$+$anomalous, as obtained by
integration with $k_0 = 0.6$~GeV and the 
$\pTmin^{\mrm{anom}}(s)$ of eq.~(\protect\ref{pTminanom}). 
By comparison, full curve is the parameterization of 
eq.~(\ref{sigtotgap}).
\label{fig7}}
\end{figure}

\begin{figure}[p]
\begin{center}
\mbox{\epsfig{file=gagafig8.ps,bbllx=0,bblly=0,bburx=360,bbury=200}}
\end{center}
\vspace{-2mm}
\caption%
{(a) Schematic graph for a hard $\gamma\p$ process, illustrating
the concept of two different scales. 
(b) The allowed phase space for this process, with the subdivision
into event classes.
\label{fig8}}
\end{figure}

\clearpage

\begin{figure}[p]
\begin{center}
\mbox{\epsfig{file=gagafig9.eps}} 
\end{center}
\vspace{-5mm} 
\caption%
{Comparison of $\gamma\gamma$ partial cross sections. Full curves:
the simple factorization ansatz of eq.~(\protect\ref{factorizedtot}). 
Dashed curves: by integration of jet cross sections (except
VMD$\times$VMD, where eq.~(\protect\ref{sigVMDgaga}) is used).
\label{fig9}}
\end{figure}

\clearpage

\begin{figure}[p]
\begin{center}
\mbox{\epsfig{file=gagafig10.eps}} 
\end{center}
\vspace{-5mm} 
\caption%
{The total $\gamma\gamma$ cross section. Full curve: the 
parameterization of eq.~(\protect\ref{sigtotgaga}). 
Dashed curve: result from sum of integrations of the six 
components.
\label{fig10}}
\end{figure}

\begin{figure}[p]
\begin{center}
\mbox{\epsfig{file=gagafig11.eps}} 
\end{center}
\vspace{-5mm} 
\caption%
{Elastic cross sections: $\rho^0\rho^0$ full curve, $\rho^0\omega$
dotted, $\rho^0\phi$ dashed and $\rho^0\Jpsi$ dash-dotted.
The latter three cross sections include both mirror-symmetric 
configurations, and have additionally been scaled up by factors 10, 10
and 1000, respectively, for better visibility.
\label{fig11}}
\end{figure}

\clearpage

\begin{figure}[p]
\begin{center}
\mbox{\epsfig{file=gagafig12.eps}} 
\end{center}
\vspace{-5mm}
\caption%
{The total transverse energy per event, separately normalized for each 
of the six event classes. Top frame: VMD$\times$VMD: full histogram;
VMD$\times$direct: dashed one; and VMD$\times$anomalous: dash-dotted 
one. Bottom frame: direct$\times$direct: full histogram;
direct$\times$anomalous: dashed one; and anomalous$\times$anomalous: 
dash-dotted one.     
\label{fig12}}
\end{figure}

\clearpage

\begin{figure}[p]
\begin{center}
\mbox{\epsfig{file=gagafig13.eps}} 
\end{center}
\vspace{-5mm}
\caption%
{The total transverse energy per event for different beams: 
$\gamma\gamma$: full histogram; $\gamma\p$: dashed one: and
$\p\p$: dash-dotted one.
\label{fig13}}
\end{figure}

\begin{figure}[p]
\begin{center}
\mbox{\epsfig{file=gagafig14.eps}} 
\end{center}
\vspace{-5mm}
\caption%
{Transverse energy flow as a function of pseudorapidity for 
different beams. Notation as in Fig.~\protect\ref{fig13}.
\label{fig14}}
\end{figure}

\clearpage

\begin{figure}[p]
\begin{center}
\mbox{\epsfig{file=gagafig15.eps}} 
\end{center}
\vspace{-7mm}
\caption%
{Transverse energy flow as a function of pseudorapidity for 
different beams, as in Fig.~\protect\ref{fig14} except that 
elastic and single diffractive events have been removed.  
Notation as in Fig.~\protect\ref{fig13}.
\label{fig15}}
\end{figure}

\begin{figure}[p]
\begin{center}
\mbox{\epsfig{file=gagafig16.eps}} 
\end{center}
\vspace{-7mm}
\caption%
{Charged particle inclusive $\pT$ spectra for different beams. 
Notation as in Fig.~\protect\ref{fig13}.
\label{fig16}}
\end{figure}

\clearpage

\begin{figure}[p]
\begin{center}
\mbox{\epsfig{file=gagafig17.eps}} 
\end{center}
\vspace{-5mm}
\caption%
{Rate of reconstructed jets as a function of the transverse jet energy 
for different beams. Notation as in Fig.~\protect\ref{fig13}.
\label{fig17}}
\end{figure}

\begin{figure}[p]
\begin{center}
\mbox{\epsfig{file=gagafig18.eps}} 
\end{center}
\vspace{-5mm}
\caption%
{Parton-level jet $\pT$ distributions for different beams.
The factor $1/2$ compensates for there being 2 jets per event.
Notation as in Fig.~\protect\ref{fig13}.
\label{fig18}}
\end{figure}

\clearpage

\begin{figure}[p]
\begin{center}
\mbox{\epsfig{file=gagafig19.eps}} 
\end{center}
\vspace{-5mm}
\caption%
{Pseudorapidity distribution of reconstructed jets for different beams. 
Notation as in Fig.~\protect\ref{fig13}.
\label{fig19}}
\end{figure}

\begin{figure}[p]
\begin{center}
\mbox{\epsfig{file=gagafig20.eps}} 
\end{center}
\vspace{-5mm}
\caption%
{Transverse energy flow for $E_{\mrm{cm}} = 50$~GeV as a 
function of pseudorapidity for different beams, cf.
Fig.~\protect\ref{fig15}.
Notation as in Fig.~\protect\ref{fig13}.
\label{fig20}}
\end{figure}


\end{document}